\documentclass[runningheads]{llncs}
\usepackage[T1]{fontenc}
\usepackage{lscape}
\usepackage{graphicx}
\usepackage{multirow}
\usepackage{subcaption}
\usepackage{caption}
\usepackage{placeins}
\usepackage{float}

\begin{document}
\title{Blood Glucose Level Prediction in Type 1 Diabetes Using Machine Learning}
\titlerunning{Blood Glucose Prediction for Diabetes Management Using Regression, Ensembles, Deep Learning, and Reinforcement Learning}
%
\author{Soon Jynn Chu\inst{1,*}\orcidID{0009-0009-1653-5855}\and
Nalaka Amarasiri\inst{1}\orcidID{0009-0003-1324-4063} \and
Sandesh Giri\inst{1}\orcidID{0000-0003-2492-3400} \and
Priyata Kafle \inst{2}\orcidID{0009-0007-4892-7955}}
\authorrunning{Chu et al.}
\titlerunning{Blood Glucose Prediction using ML}
%
\institute{University of Louisiana at Lafayette, Lafayette, LA 70503, USA  \and
MBBS, Hospital for Advanced Medicine \& Surgery (HAMS), Kathmandu, Nepal}
\maketitle              
\begin{abstract}

Type 1 Diabetes is a chronic autoimmune condition in which the immune system attacks and destroys insulin-producing beta cells in the pancreas, resulting in little to no insulin production. Insulin helps glucose in your blood enter your muscle, fat, and liver cells so they can use it for energy or store it for later use. If insulin is insufficient, it causes sugar to build up in the blood and leads to serious health problems. People with Type 1 Diabetes need synthetic insulin every day. In diabetes management, continuous glucose monitoring is an important feature that provides near real-time blood glucose data. It is useful in deciding the synthetic insulin dose. In this research work, we used machine learning tools, deep neural networks, deep reinforcement learning, and voting and stacking regressors to predict blood glucose levels at 30-min time intervals using the latest DiaTrend dataset. Predicting blood glucose levels is useful in better diabetes management systems. The trained models were compared using several evaluation metrics. Our evaluation results demonstrate the performance of various models across different glycemic conditions for blood glucose prediction. The source codes of this work can be found in:

\verb|https://github.com/soon-jynn-chu/t1d_bg_prediction|

\end{abstract}

\section{Introduction}

Type 1 Diabetes (T1D) is a chronic autoimmune condition characterized by the destruction of insulin-producing beta cells in the pancreas, resulting in little to no endogenous insulin production \cite{atkinson2014type}. This deficiency leads to dysregulation of blood glucose levels, necessitating lifelong external insulin administration to maintain normoglycemia \cite{tridgell2010inpatient}. T1D affects millions of people worldwide, with its incidence increasing by approximately 3-4\% annually, particularly in younger populations \cite{maahs2010epidemiology}. The management of T1D is complex and demanding, requiring constant vigilance to balance insulin dosing with various factors that influence blood glucose levels, such as food intake, physical activity, stress, and illness \cite{holt2021management}. The primary goal of T1D management is to maintain blood glucose levels within a target range to prevent both short-term complications and long-term microvascular and macrovascular complications \cite{livingstone2020tightly}.

Understanding the various glycemic values is crucial for effective T1D management. Blood glucose levels are categorized as hypoglycemia when below 70 mg/dL (3.9 mmol/L), within the target range of 70-180 mg/dL (3.9-10.0 mmol/L), and hyperglycemia when above 180 mg/dL (10.0 mmol/L). Hemoglobin A1C (HbA1c), which indicates average blood glucose levels over the past 2-3 months, has a target of less than 7.0\% for most adults with T1D. Time In Range (TIR) is another important metric, intending to spend more than 70\% of time within the target range \cite{care2023standards}. Glycemic variability, measured by metrics such as standard deviation or the coefficient of variation, is also significant, with lower variability associated with better outcomes \cite{mo2024glycemic}.

In recent years, technological advancements have revolutionized diabetes management, particularly through the development of Continuous Glucose Monitoring (CGM) systems. CGM devices are wearable sensors that measure interstitial glucose levels at regular intervals, typically every 5 minutes, providing near real-time glucose data to the user \cite{mihai2022continuous}. This continuous stream of information offers several advantages over traditional fingerstick blood glucose measurements, including improved glycemic control through more frequent glucose readings, early detection and prevention of hypo- and hyperglycemic events, identification of glucose trends and patterns, and enhanced decision-making for insulin dosing and lifestyle choices \cite{edelman2018clinical}.

CGM systems provide a wealth of data that can be interpreted to improve diabetes management. The current glucose reading displays the most recent interstitial glucose measurement, updated every 5 minutes for most systems \cite{matabuena2023reproducibility}. Trend arrows indicate the direction and rate of glucose change, helping predict short-term glucose fluctuations. The Ambulatory Glucose Profile (AGP) offers a visualization of glucose data over time, aiding in the identification of patterns and trends \cite{miller2020using}. The Glucose Management Indicator (GMI) estimates HbA1c based on average CGM glucose readings, providing a more immediate indicator of glycemic control. Additionally, customizable alerts and alarms for high and low glucose levels, including predictive alerts for impending hypo- or hyperglycemia, enhance the user's ability to manage their condition proactively \cite{friedman2023beyond}.


In the context of advancing diabetes management technologies, the DiaTrend \cite{prioleau2023diatrend} dataset has emerged as an invaluable resource for researchers and data scientists. This comprehensive dataset comprises intensive longitudinal data from wearable medical devices, including 27,561 days of CGM data and 8,220 days of insulin pump data from 54 patients with T1D, aged 19-74 years, with 17 males and 37 females. The richness and scope of the DiaTrend dataset provide an unprecedented opportunity to develop and validate Machine Learning (ML) models for various aspects of diabetes care, including blood glucose prediction, detection of adverse glycemic events, optimization of insulin delivery, discovery of digital biomarkers for glycemic control, and analysis of adherence patterns to wearable medical devices.

Our research leverages the DiaTrend dataset to develop and compare various ML methods for predicting blood glucose levels using CGM data in 30-minute intervals. We explore a range of algorithms, including Support Vector Regressor (SVR), Light Gradient Boosting Machine (LGB), Multi-Layer Perceptron (MLP), Long Short-Term Memory (LSTM) networks, Gated Recurrent Units (GRU), Deep Deterministic Policy Gradient (DDPG), Twin Delayed Deep Deterministic Policy Gradient (TD3), and Soft Actor-Critic (SAC). By evaluating these diverse approaches, we aim to identify the most effective methods for accurate insulin level prediction. This research has the potential to significantly enhance diabetes management strategies and contribute to the development of more sophisticated closed-loop systems.

The potential impact of this research extends beyond immediate improvements in glycemic control. Accurate prediction of blood glucose levels can lead to more precise insulin dosing recommendations, reduced risk of hypoglycemia and hyperglycemia, improved overall quality of life for individuals with T1D, advancements in personalized medicine approaches for diabetes management, and development of more robust artificial pancreas systems. Furthermore, the insights gained from this research could inform the design of next-generation CGM systems and insulin pumps, potentially leading to enhanced prediction algorithms integrated directly into diabetes management devices, improved decision support tools for healthcare providers and patients, and more effective strategies for managing challenging scenarios such as exercise, illness, and stress.

\section{Related Work}

Recent advances in ML and Artificial Intelligence (AI) have spurred the development of innovative models for predicting blood glucose levels, optimizing insulin delivery, and improving diabetes management, particularly for individuals with T1D. The related works discussed in this section address various aspects of T1D care, including adverse event prediction, insulin dosing, and personalized treatment approaches, leveraging both novel datasets and advanced modeling techniques.

Zheng et al. \cite{zheng2024predicting} proposed BG-BERT, a self-supervised learning framework designed to predict blood glucose levels with a specific focus on adverse events such as hyperglycemia and hypoglycemia, which are often underrepresented in datasets. BG-BERT employs a masked autoencoder to capture contextual information from blood glucose records and uses the Synthetic Minority Oversampling Technique (SMOTE), a data augmentation technique, to enhance sensitivity to adverse events. Evaluated against benchmark datasets, BG-BERT demonstrated a 9.5\% improvement in prediction accuracy and a 44.9\% increase in sensitivity to adverse events compared to state-of-the-art models. This work underscores the importance of incorporating novel data augmentation techniques and self-supervised learning to address the imbalance inherent in adverse event prediction in T1D datasets.

Rodriguez-Leon et al. \cite{rodriguez2023t1diabetesgranada} introduced the T1DiabetesGranada dataset, one of the largest open longitudinal datasets for T1D research, comprising 257,780 days of continuous glucose monitoring (CGM) data from 736 patients over four years. This dataset, which includes demographic and clinical data, addresses the common issue of data scarcity in diabetes research, facilitating the development of more robust AI models for glucose level prediction and management. The dataset’s multimodal nature allows researchers to integrate various data types, such as glucose levels and patient demographics, enabling more comprehensive models for disease management.

Lu et al. \cite{lu2024mealtime} focused on predicting mealtime patterns using insulin pump data to improve postprandial glucose control in T1D patients. Their research highlighted the irregularity of meal timing among patients and developed personalized LSTM-based models that achieved high accuracy (F1 score > 95\%) in predicting future mealtimes. This mealtime prediction framework could significantly enhance the timing of insulin dosing, reducing post-meal hyperglycemia. The ability to predict meals represents a crucial step in automating insulin delivery systems and improving patient adherence to recommended insulin dosing schedules.

Kurdi et al. \cite{kurdi2023proof} explored the use of Supervised Machine Learning Algorithms (SMLAs) to predict glycemic control and adherence to Insulin Pump Self-Management Behaviors (IPSMB). Their study compared three algorithms (logistic regression, random forest, and k-nearest neighbor) demonstrating that baseline hemoglobin A1c, continuous glucose monitoring usage, and patient sex were predictive of adherence to IPSMB criteria. The models achieved comparable predictive performance, with random forest showing superior calibration. This work highlights the potential of SMLAs for predicting patient behaviors and improving long-term glycemic outcomes.

Reinforcement Learning (RL) has been successfully applied in various domains, including robotics and microgrid control applications, which demonstrates its versatile capability. \cite{amarasiri2024universal},\cite{sheida2024adaptive},\cite{sheida2024resilient}. Moreover, multimodal learning and attention-based frameworks have shown significant promise in healthcare applications by leveraging complementary data sources and adaptive mechanisms to improve model performance and robustness \cite{bodaghi2024multimodal},\cite{bodaghi2024adaptive},\cite{hosseini2023multimodal}.

Zhu et al. \cite{zhu2020basal} and Mohammad Javad et al. \cite{javad2019reinforcement} employed RL to tackle the complexity of glucose regulation and insulin dosing in T1D patients. Zhu et al. applied Deep Reinforcement Learning (DRL) to develop basal glucose control strategies using both single-hormone (insulin) and dual-hormone (insulin and glucagon) delivery systems. Their work demonstrated significant improvements in time-in-range for glucose levels, particularly in adolescent cohorts, showcasing the viability of RL for closed-loop glucose control. Similarly, Mohammad Javad et al. developed a Q-learning-based RL algorithm for personalized insulin dosage recommendations. The algorithm adapted to patient-specific characteristics, such as glycated hemoglobin levels, body mass index, and physical activity, to suggest appropriate insulin doses. This approach achieved an 88\% match with physician-prescribed insulin doses, demonstrating the potential for RL to optimize personalized insulin therapy.

These studies collectively highlight the diverse approaches being taken to enhance T1D management through AI and ML. From leveraging large-scale datasets like DiaTrend and T1DiabetesGranada to employing advanced techniques such as self-supervised learning, RL, and personalized predictive models, these works contribute to improving glycemic control, reducing adverse events, and advancing the development of closed-loop insulin delivery systems. Our research builds on these advances by using CGM data to develop ML models that predict blood glucose levels, aiming to contribute to the growing body of work focused on enhancing diabetes care through technology-driven solutions.

\section{Materials \& Methods}
\subsection{Data Acquisition}
The DiaTrend \cite{prioleau_diatrend_2023} dataset was utilized for this study. It consists of more than 27 thousand days of CGM data and more than 8 thousand days of insulin pump data collected from 54 patients with T1D. It is a combination of two independent studies: Digital SMD and SweetGoals. In this research, we utilize only the CGM features and data from the first study, which consisted of 17 patients between 25 and 74 years of age. The purpose was to reduce the computational time.  The blood glucose levels were sampled at an average of five-minute intervals using FDA-approved CGMs. 

\begin{figure}[htbp]
    \centering
    \begin{subfigure}[b]{0.49\textwidth}
        \centering
        \includegraphics[width=\textwidth]{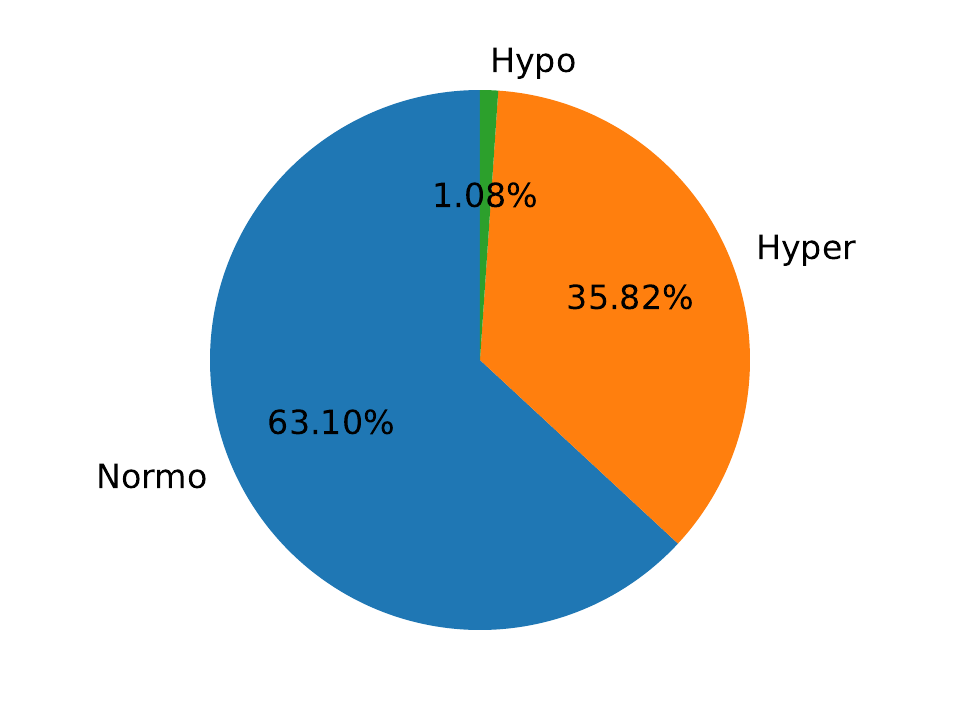}
        \caption{Train Set}
        \label{fig:train_dist}
    \end{subfigure}
    \hfill
    \begin{subfigure}[b]{0.49\textwidth}
        \centering
        \includegraphics[width=\textwidth]{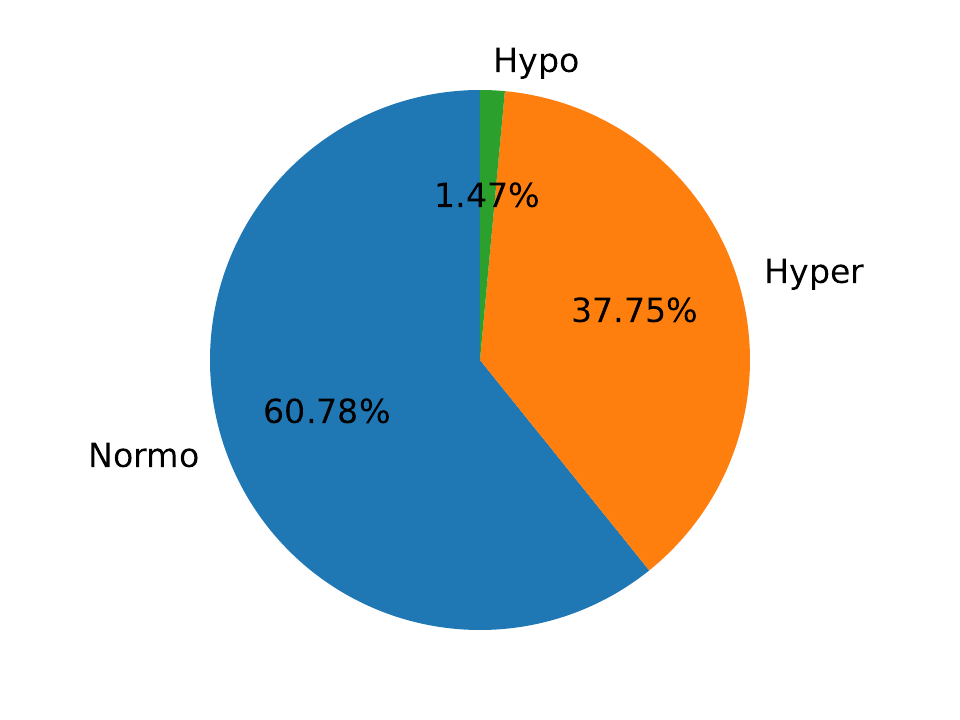}
        \caption{Test Set}
        \label{fig:test_dist}
    \end{subfigure}
    \caption{Distribution for Each Condition for Train and Test Set}
    \label{fig:condition_dist}
\end{figure}

We randomly split 11 subjects (S30, S31, S38, S39, S46, S47, S49, S50, S52, S53, S54) and 6 subjects (S29, S36, S37, S42, S45, S51) for the training and test set. For a prediction horizon of 30 minutes, we used one hour of data to create a sequence of samples for a supervised machine-learning task. The initial 30 minutes of data were used as the input sequence, while the subsequent 30 minutes were used as the output sequence. Both the input and output sequences consisted of 6 data points. We removed the samples with missing values because we have sufficient data for training and testing. 36.52\% of the data shows hyperglycemia, while 1.22\% shows hypoglycemia. We have 250,559 samples, with 159,557 for training and 91,002 for testing. The distribution for each glycemic condition for the train and test set is shown in Figure \ref{fig:condition_dist}. They consist of approximately 62\% normoglycemia, around 37\% hyperglycemia, and about 1\% hypoglycemia.


The data was normalized between -1 and 1 using Eq.~\ref{eqn:normalizing}, where $\min{x}$ and $\max{x}$ were set to constant values of 20 and 420. 

\begin{equation}
\label{eqn:normalizing}
    Normalized~data = [2 * \frac{x - \min{x}}{\max{x} - \min{x}} - 1]
\end{equation}

\subsection{Blood Glucose Prediction Models}

\subsubsection{Traditional Machine Learning (ML)}
We examine the performance of three traditional ML algorithms: Support Vector Regression (SVR), Random Forest (RF), and Light Gradient-Boosting Machine (LGB).

SVR aims to find a hyperplane that maximizes the margin between predicted and actual values. RF combines predictions from multiple decision trees, each trained on a random subset of the data, to produce a final prediction. LGB is a gradient-boosting framework that builds upon decision trees and employs gradient descent and boosting techniques to minimize a specified cost function.

Since traditional ML models typically don't support multi-output prediction, we used the last value of the output sequence as the target for prediction. We kept all hyperparameters at default values to ensure a fair model comparison during training. The SVR and RF models were implemented using the \verb|scikit-learn| (v1.5.2) library.

\subsubsection{Deep Neural Networks (DNN)}
DNNs are class of ML that try to mimic the function of a human brain. It consists of multiple layers of interconnected nodes, allowing it to learn complex patterns in the data. In this study, we investigate the performance of three different DNN architectures: Multi-Layer Perception (MLP), Long Short-Term Memory (LSTM), and Gated Recurrent Unit (GRU). 

MLP is a fundamental architecture of a DNN. It consists of interconnected layers of nodes with nonlinear activation functions. LSTM and GRU are types of Recurrent Neural Networks (RNNs). RNNs process data sequentially, maintaining a hidden state that captures information from previous inputs to determine the current input and output. LSTM addresses the vanishing gradient problem in RNNs by introducing three gates (input, forget, and output) as cells in the hidden state. On the other hand, GRU addresses the short-term memory of RNNs by introducing two gates (reset and update) in the hidden state. Both architectures aim to regulate the flow of information in the network to achieve the desired output. 

\begin{figure}[htbp]
    \centering
    \begin{subfigure}{\textwidth}
        \centering
        \includegraphics[width=\textwidth]{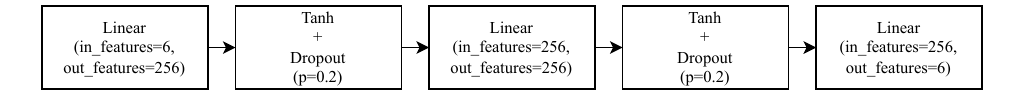}
        \caption{MLP Architecture}
        \label{fig:mlp_architecture}
    \end{subfigure}
    
    \begin{subfigure}{\textwidth}
        \centering
        \includegraphics[width=0.433\textwidth]{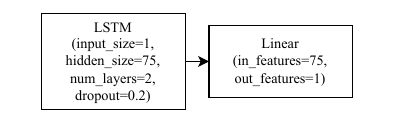}
        \caption{LSTM Architecture}
        \label{fig:lstm_architecture}
    \end{subfigure}

    \begin{subfigure}{\textwidth}
        \centering
        \includegraphics[width=0.433\textwidth]{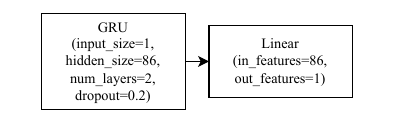}
        \caption{GRU Architecture}
        \label{fig:gru_architecture}
    \end{subfigure}
    
    \caption{DNN Model Architecture Illustration}
    \label{fig:dnn-model-architectures}
\end{figure}

The DNN models described in this paper, as shown in Figure \ref{fig:dnn-model-architectures}, take the entire input sequence of CGM data and generate an output sequence of the same length. The MLP model, as shown in Figure \ref{fig:mlp_architecture}, comprised two fully connected layers with a hidden size of 256. It uses a hyperbolic tangent (Tanh) activation function with a dropout rate of 20\% between layers. This results in 69,126 parameters in the network. The LSTM model, as shown in Figure \ref{fig:lstm_architecture}, consists of two layers with a hidden size of 75 and a dropout rate of 20\%. It is followed by a fully connected layer, resulting in 69,076 parameters. As shown in Figure \ref{fig:gru_architecture}, the GRU model was implemented similarly to the LSTM model but with an 86 hidden size, resulting in 67,941 parameters.

All DNNs were implemented and trained using \verb|PyTorch| (v2.4.0). The training hyperparameters are shown in Table \ref{tab:dnn-hyperparameters}. The cost function was the Huber loss, with the Adam optimizer set at a 0.01 learning rate. The models were trained for a maximum of 100 epochs, with a batch size of 32. Learning rate scheduling and early stopping were introduced to prevent overfitting while training. The scheduler reduces the learning rate by a factor of 0.1 when the model does not show improvement for three consecutive epochs, with a minimum cap of 1e-6. Training was stopped when the model did not show improvement for five epochs in a row.

\begin{table}[htbp]
\centering
\caption{Summary of Hyperparameters Used in DNN Model Training}
\label{tab:dnn-hyperparameters}
\begin{tabular}{cc}
\hline
Hyperparameter & Value \\ \hline
Criterion & Huber Loss \\
Optimizer & Adam \\
Epoch & 100 \\
Batch Size & 32 \\
Initial Learning Rate & 0.01 \\
Learning Rate Reduction Factor & 0.1 \\ 
Learning Rate Scheduler Patience & 3 \\ 
Minimum Learning Rate & 1e-6 \\ 
Early Stopping Patience & 5 \\ \hline
\end{tabular}
\end{table}

\subsubsection{Deep Reinforcement Learning (DRL)}
Reinforcement Learning (RL) is a branch of machine learning that trains an agent to make decisions in an environment. The agent learns through trial and error and receives a reward signal as feedback from the environment. Our primary focus is on variants of Q-learning, a model-free RL algorithm that learns a policy that maps states to actions and evaluates the quality of each action at a given state. Q-learning aims to learn an action-value function, the Q-function, which quantifies the expected return of taking a specific action in a particular state. When the Q-function is represented using deep neural networks, it becomes a DRL algorithm, allowing for more complex and high-dimensional state spaces.

In this paper, the goal of the agent is to predict the output sequence given the input sequence of the CGM data. We selected the Normalized Root Mean Squared Error (NRMSE), as shown in Equation \ref{eqn:reward_function}, as the reward function because it provides bounded and scaled values between 0 and 1.

\begin{equation}
\label{eqn:reward_function}
    NRMSE = 1 - \frac{RMSE}{\max y - \min y}
\end{equation}

We experimented with three Q-learning methods for continuous action space: Deep Deterministic Policy Gradient (DDPG) \cite{lillicrap_continuous_2019}, Twin Delayed DDPG (TD3) \cite{fujimoto_addressing_2018}, and Soft Actor-Critic (SAC) \cite{haarnoja_soft_2018}. DDPG utilizes the Bellman equation to learn the Q-function and then leverages the Q-function to learn the policy. TD3 improves upon DDPG by training two Q-functions and preventing Q-function exploitation through action noise addition. SAC learns a stochastic policy that connects with DDPG, which is taught to optimize the balance between the expected return and randomness in the policy.

All three DRL algorithms utilize the same Q-function architecture, consisting of an MLP with two fully connected layers with a hidden size of 256. Additionally, a 20\% dropout rate and the Tanh activation function are employed. They were implemented using the same hyperparameters in the \verb|d3rlpy| (v2.6.0) library \cite{seno_d3rlpy_2022}. They underwent 10 training epochs with a 1e-4 learning rate, while the other hyperparameters remained at their default settings. 

\subsubsection{Voting \& Stacking Regressor}
The voting regressor is an ensemble method that averages the outputs of several base estimators. On the other hand, the stacking regressor is an ensemble method that stacks the prediction of several base estimators and trains a final estimator to obtain the final output. These models aim to enhance predictive performance by utilizing the strengths of multiple algorithms, thereby reducing the risk of overfitting and improving generalizability. In this paper, we experimented with three voting and stacking regressors with the different base estimators as shown in Table \ref{tab:vote-stack-configurations}. The base estimators of both types were combined based on their categories: ML, DNN, or DRL. We chose Simple Linear Regression (SLR) as the final estimator for all stacking models as it provides a straightforward and interpretable approach.

\begin{table}[htbp]
\centering
\caption{Model Configurations for Voting and Stacking Regressors}
\label{tab:vote-stack-configurations}
\renewcommand{\arraystretch}{1.25}
\begin{tabular}{cccc}
\hline
Type & Model Name & Base Estimators & Final Estimator \\ \hline
\multirow{3}{*}{Voting} & V1 & SVR, RF, LightGBM & - \\
 & V2 & MLP, LSTM, GRU & - \\
 & V3 & SAC, DDPG, TD3 & - \\ \hline
\multirow{3}{*}{Stacking} & S1 & SVR, RF, LightGBM & SLR \\
 & S2 & MLP, LSTM, GRU & SLR \\
 & S3 & SAC, DDPG, TD3 & SLR \\ \hline
\end{tabular}
\end{table}

\subsection{Evaluation Metrics}
The metrics used to evaluate each method are listed below:

\begin{itemize}
    \item \textbf{Mean absolute error (MAE)}: MAE calculates the average magnitude of the absolute difference between the predicted value $\hat{y}_i$ and actual value $y_i$.
        \begin{equation}
            MAE = \frac{1}{N}\sum_{i=1}^n|\hat{y}_i - y_i|
        \end{equation}
    \item \textbf{Mean absolute percentage error (MAPE)}: MAPE expresses MAE as a percentage.
        \begin{equation}
            MAPE = \frac{1}{N}\sum_{i=1}^n\frac{|\hat{y}_i - y_i|}{y_i}
        \end{equation}
    \item \textbf{Root mean squared error (RMSE)}: RMSE calculates the square root of the average squared difference between $\hat{y}_i$ and $y_i$.
        \begin{equation}
            RMSE = \sqrt{\frac{\sum_{i=1}^N(\hat{y}_i - y_i)^2}{N}}
        \end{equation}
\end{itemize}

\section{Results}
We assessed the performance of various models using the DiaTrend dataset with a 30-minute prediction horizon. The models were trained on 11 subjects and then evaluated on the 6 remaining subjects. We used MAE, MAPE, and RMSE as the evaluation metrics, with RMSE being the deciding factor. The overall results are presented in Table \ref{tab:cumulative-performance}, showing the average metrics across all 6 subjects along with the standard deviation. The results are presented in actual scale. Model S2 demonstrated the best performance in terms of MAE and MAPE, with values of 16.29 and 10.40\%, respectively. However, Model V2 achieved an RMSE of 22.50, outperforming Model S2 by a magnitude of 0.03, making it the best-performing model overall.

In Table \ref{tab:glycemic-performance}, we compare the performance of different models under varying glycemic conditions. In the normoglycemic condition, LSTM and Model S2 delivered the top performance with an RMSE of 19.03. Model V2 excelled in the hyperglycemic region with an RMSE of 26.02. In the hypoglycemic condition, DDPG achieved the lowest RMSE of 28.25, outperforming the second-best model, MLP, by 0.79 units.


\begingroup
\setlength{\tabcolsep}{12pt}
\begin{table}[htbp]
\centering
\caption{Cumulative Performance of Various Models (Mean and Standard Deviation)}
\label{tab:cumulative-performance}
\renewcommand{\arraystretch}{1.6}
\begin{tabular}{cccc}
\hline
Model & MAE                  & MAPE (\%)            & RMSE                 \\ \hline
SVR   & 16.80(2.42)          & 10.81(1.87)          & 22.93(3.24)          \\
RF    & 17.26(2.53)          & 11.01(1.86)          & 23.63(3.33)          \\
LGB   & 16.81(2.58)          & 10.75(1.91)          & 23.15(3.45)          \\ \hline
MLP   & 16.67(2.53)          & 10.54(1.91)          & 22.84(3.38)          \\
LSTM  & 16.42(2.42)          & 10.62(1.87)          & 22.59(3.22)          \\
GRU   & 16.38(2.52)          & 10.51(1.91)          & 22.57(3.36)          \\ \hline
DDPG  & 17.78(2.97)          & 10.99(2.15)          & 24.75(4.00)          \\
TD3   & 18.58(3.14)          & 11.69(2.27)          & 25.44(4.18)          \\
SAC   & 18.52(2.96)          & 11.34(2.04)          & 25.48(3.91)          \\ \hline
V1    & 16.63(2.52)          & 10.65(1.87)          & 22.85(3.36)          \\
V2    & 16.38(2.49)          & 10.47(1.88)          & \textbf{22.50(3.33)} \\
V3    & 17.89(3.06)          & 11.14(2.19)          & 24.78(4.04)          \\  \hline
S1    & 16.56(2.54)          & 10.53(1.85)          & 22.84(3.39)          \\
S2    & \textbf{16.29(2.50)} & \textbf{10.40(1.86)} & 22.53(3.34)          \\
S3    & 17.84(2.89)          & 11.28(2.15)          & 24.42(3.88)          \\ \hline

\end{tabular}
\end{table}
\endgroup

\begin{landscape}
\begin{table}[htbp]
\centering
\caption{Performance for Various Models Across Different Glycemic Conditions (Mean and Standard Deviation)}
\label{tab:glycemic-performance}
\renewcommand{\arraystretch}{1.75}
\begin{tabular}{cccccccccc}
\hline
\multirow{2}{*}{Model} & \multicolumn{3}{c}{Normoglycemia} & \multicolumn{3}{c}{Hyperglycemia} & \multicolumn{3}{c}{Hypoglycemia} \\ \cline{2-10} 
 & MAE & MAPE (\%) & RMSE & MAE & MAPE (\%) & RMSE & MAE & MAPE (\%) & RMSE \\ \hline
SVR & 14.61(1.97) & 11.32(1.62) & 19.54(2.88) & 19.64(3.60) & 8.65(1.70) & 26.71(4.19) & 31.47(3.00) & 54.20(5.28) & 34.99(3.95) \\
RF & 14.93(2.04) & 11.43(1.63) & 20.23(2.86) & 20.33(3.65) & 8.97(1.72) & 27.41(4.31) & 33.54(5.27) & 58.06(9.22) & 37.88(5.56) \\
LGB & 14.32(2.14) & 11.01(1.74) & 19.46(3.09) & 20.06(3.76) & 8.84(1.78) & 27.17(4.42) & 34.41(4.10) & 59.54(7.22) & 37.71(4.66) \\ \hline
MLP & 14.85(2.19) & 11.36(1.77) & 20.12(3.17) & 19.18(3.40) & \textbf{8.41(1.61)} & 26.05(4.04) & 22.85(4.05) & 39.33(7.08) & 29.04(5.43) \\
LSTM & 14.11(1.96) & 10.95(1.59) & \textbf{19.03(2.86)} & 19.30(3.55) & 8.50(1.67) & 26.31(4.15) & 35.78(3.87) & 61.64(6.74) & 38.84(4.19) \\
GRU & 14.24(2.10) & 10.94(1.69) & 19.35(3.05) & \textbf{19.10(3.56)} & 8.42(1.68) & 26.07(4.19) & 32.55(4.17) & 56.44(7.58) & 36.46(5.14) \\ \hline
DDPG & 15.16(2.70) & 11.41(2.13) & 21.07(3.78) & 21.50(3.97) & 9.47(1.91) & 29.10(4.85) & \textbf{22.02(4.00)} & \textbf{38.30(7.05)} & \textbf{28.25(6.21)} \\
TD3 & 16.82(2.60) & 12.55(2.08) & 22.85(3.74) & 21.17(4.28) & 9.46(2.03) & 28.75(5.11) & 25.46(4.02) & 44.14(6.94) & 31.17(5.82) \\
SAC & 15.26(2.47) & 11.46(1.99) & 21.03(3.46) & 23.32(4.15) & 10.24(1.98) & 30.89(4.88) & 25.08(4.00) & 43.57(6.86) & 30.47(5.52) \\ \hline
V1 & 14.27(2.07) & 10.99(1.66) & 19.33(2.98) & 19.71(3.68) & 8.69(1.73) & 26.74(4.32) & 33.07(4.10) & 57.16(7.22) & 36.49(4.72) \\
V2 & 14.28(2.08) & 10.97(1.67) & 19.35(3.02) & 19.11(3.51) & \textbf{8.41(1.66)} & \textbf{26.02(4.12)} & 30.00(3.93) & 51.86(6.94) & 34.07(4.81) \\
V3 & 15.54(2.60) & 11.67(2.07) & 21.40(3.66) & 21.32(4.23) & 9.42(2.01) & 28.94(4.99) & 24.12(3.90) & 41.91(6.78) & 29.86(5.73) \\ \hline
S1 & 14.16(2.07) & 10.85(1.66) & 19.27(2.98) & 19.78(3.71) & 8.74(1.75) & 26.89(4.33) & 30.84(3.74) & 53.31(6.56) & 34.44(4.51) \\
S2 & \textbf{13.94(2.06)} & \textbf{10.69(1.64)} & \textbf{19.03(2.97)} & 19.36(3.58) & 8.54(1.69) & 26.39(4.20) & 32.23(4.01) & 55.75(7.07) & 35.79(4.58) \\
S3 & 15.60(2.51) & 11.92(2.02) & 21.10(3.63) & 20.87(4.00) & 9.18(1.90) & 28.26(4.77) & 27.14(3.80) & 46.98(6.69) & 32.36(5.64) \\ \hline
\end{tabular}
\end{table}
\end{landscape}

\begin{figure}[H]
    \centering
    \begin{minipage}[t]{0.48\textwidth}
        \centering
        \includegraphics[width=\textwidth]{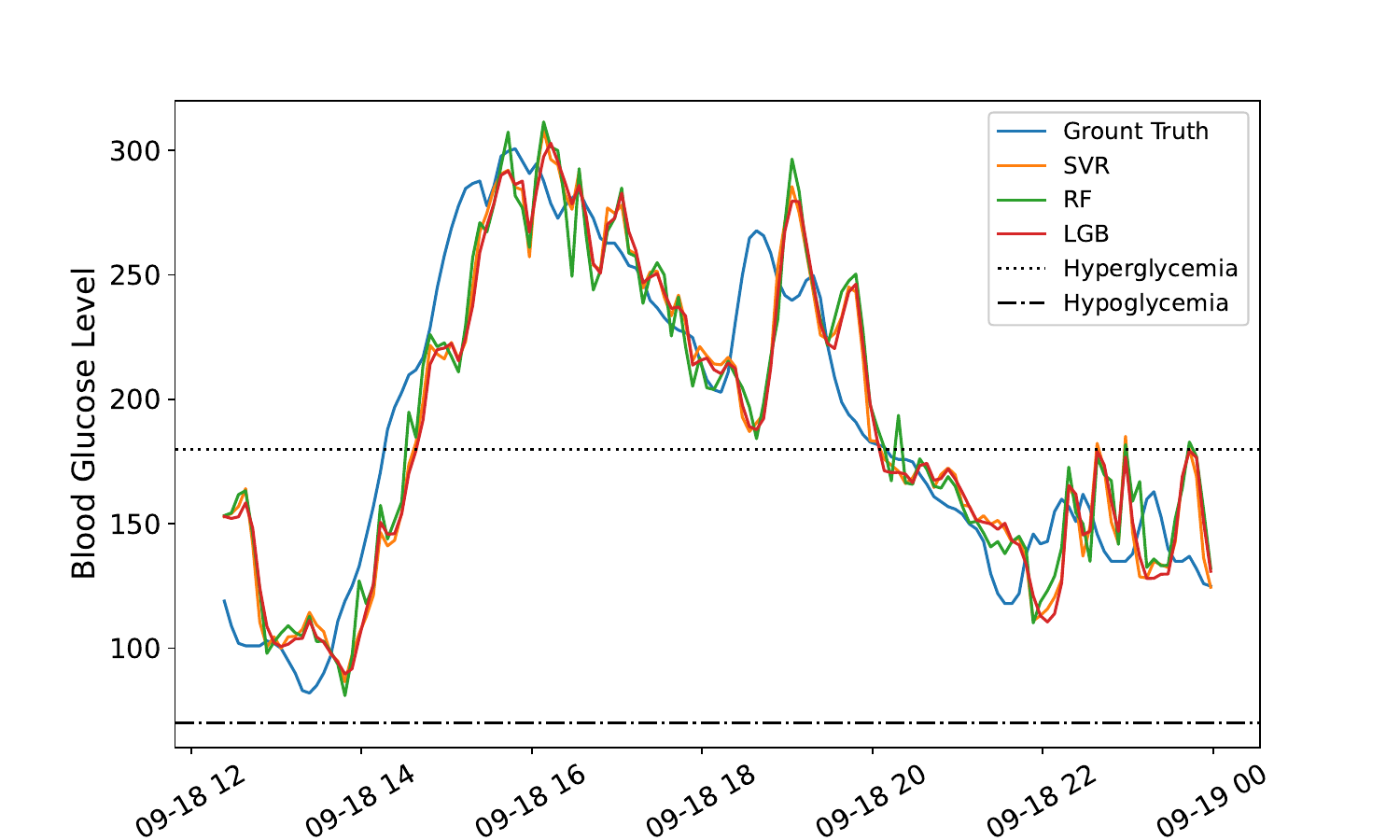}
        \caption{ML methods prediction visualization}
        \label{fig:ml_plot}
    \end{minipage}
    \hfill
    \begin{minipage}[t]{0.48\textwidth}
        \centering
        \includegraphics[width=\textwidth]{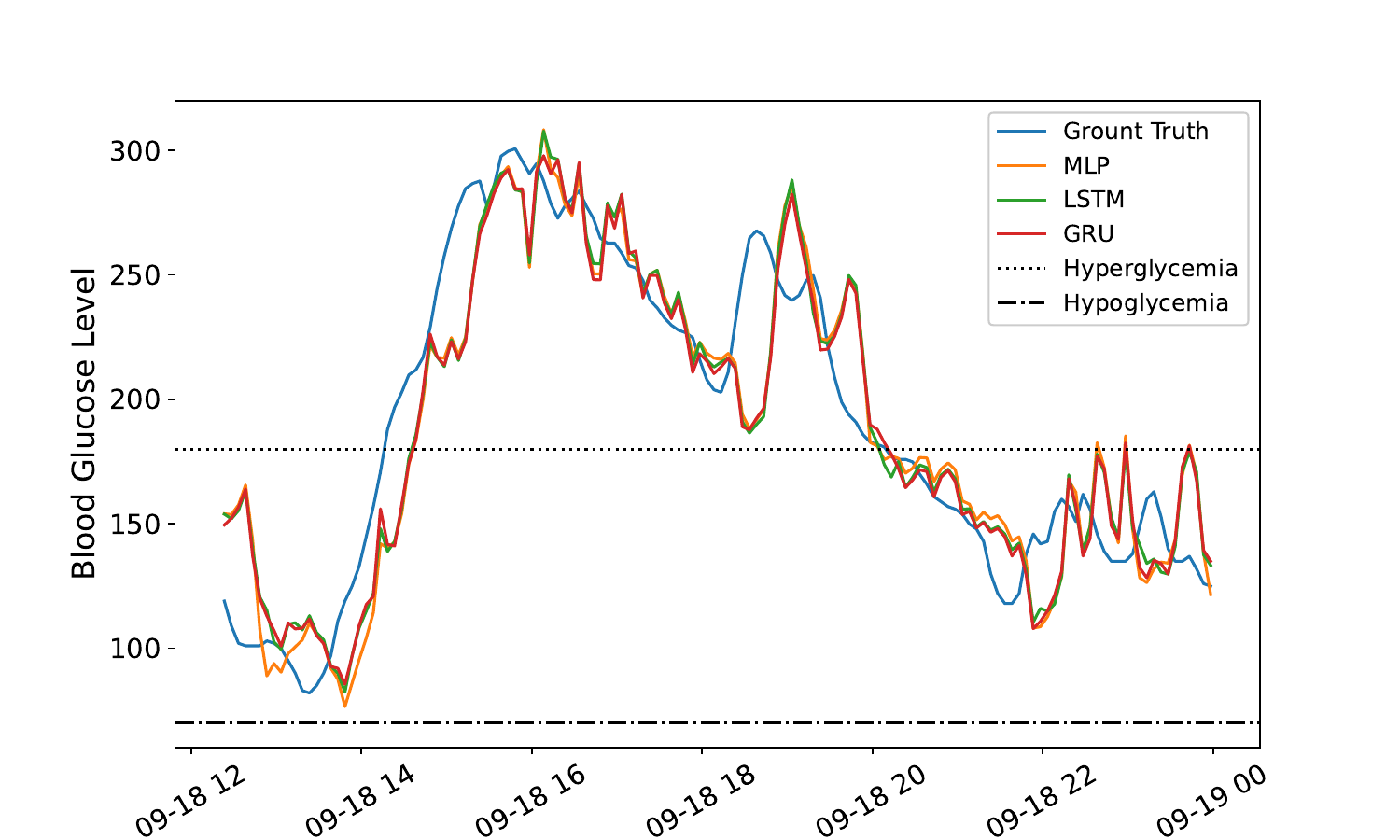}
        \caption{NN methods prediction visualization}
        \label{fig:nn_plot}
    \end{minipage}

    \vspace{5pt} 

    \begin{minipage}[t]{0.48\textwidth}
        \centering
        \includegraphics[width=\textwidth]{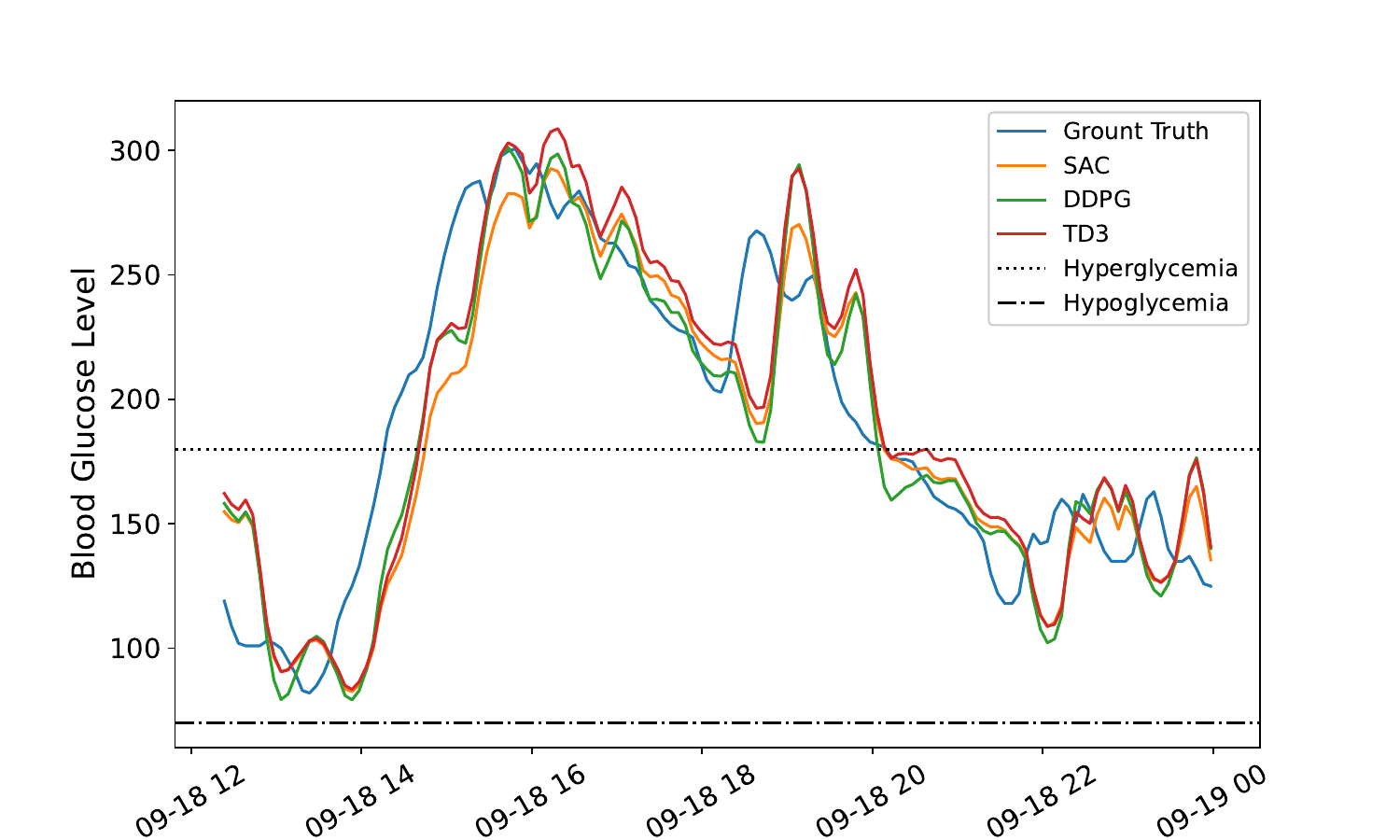}
        \caption{RL methods prediction visualization}
        \label{fig:rl_plot}
    \end{minipage}
    \hfill
    \begin{minipage}[t]{0.48\textwidth}
        \centering
        \includegraphics[width=\textwidth]{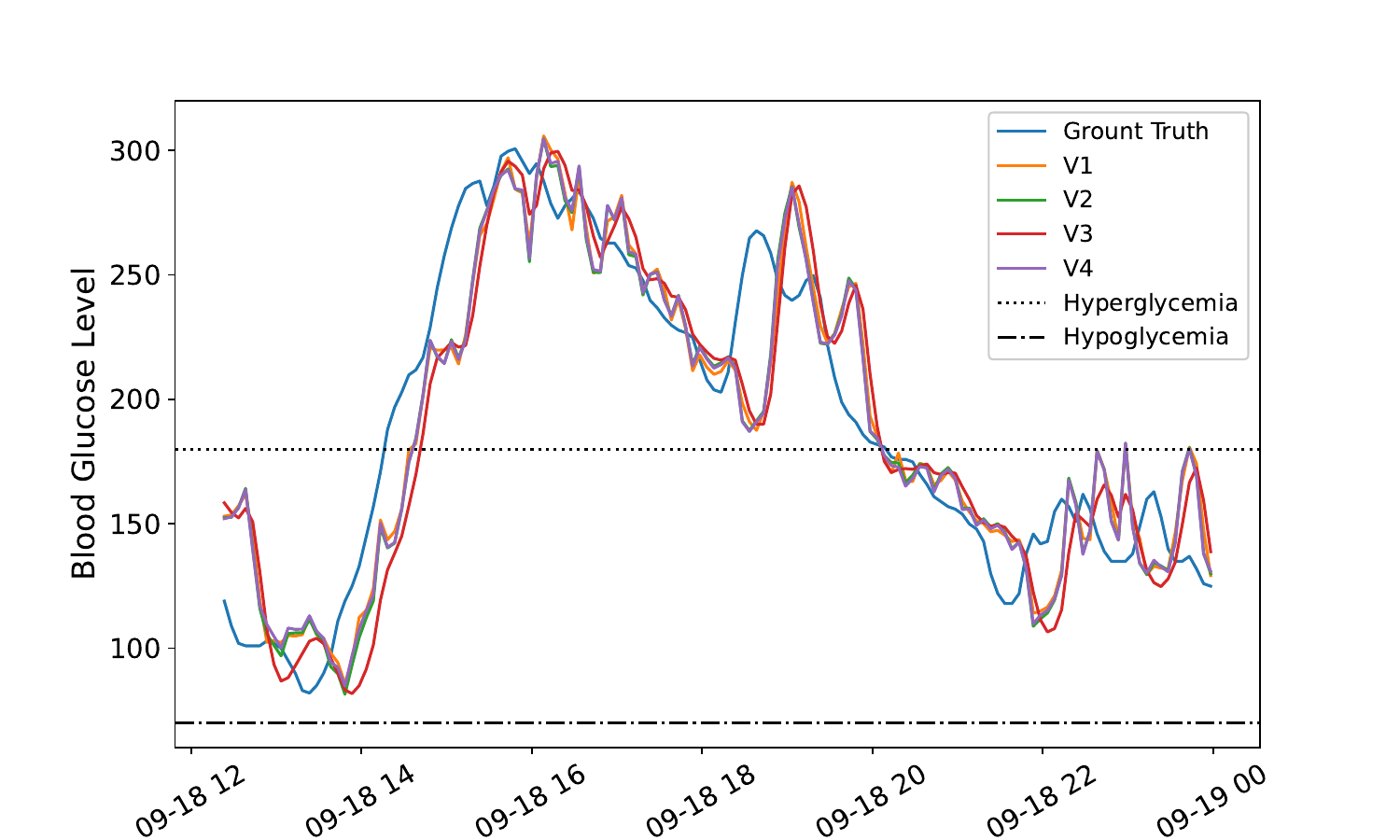}
        \caption{Voting regressor methods prediction visualization}
        \label{fig:vote_plot}
    \end{minipage}

    \vspace{5pt} 

    \begin{minipage}[t]{0.65\textwidth}
        \centering
        \includegraphics[width=0.75\textwidth]{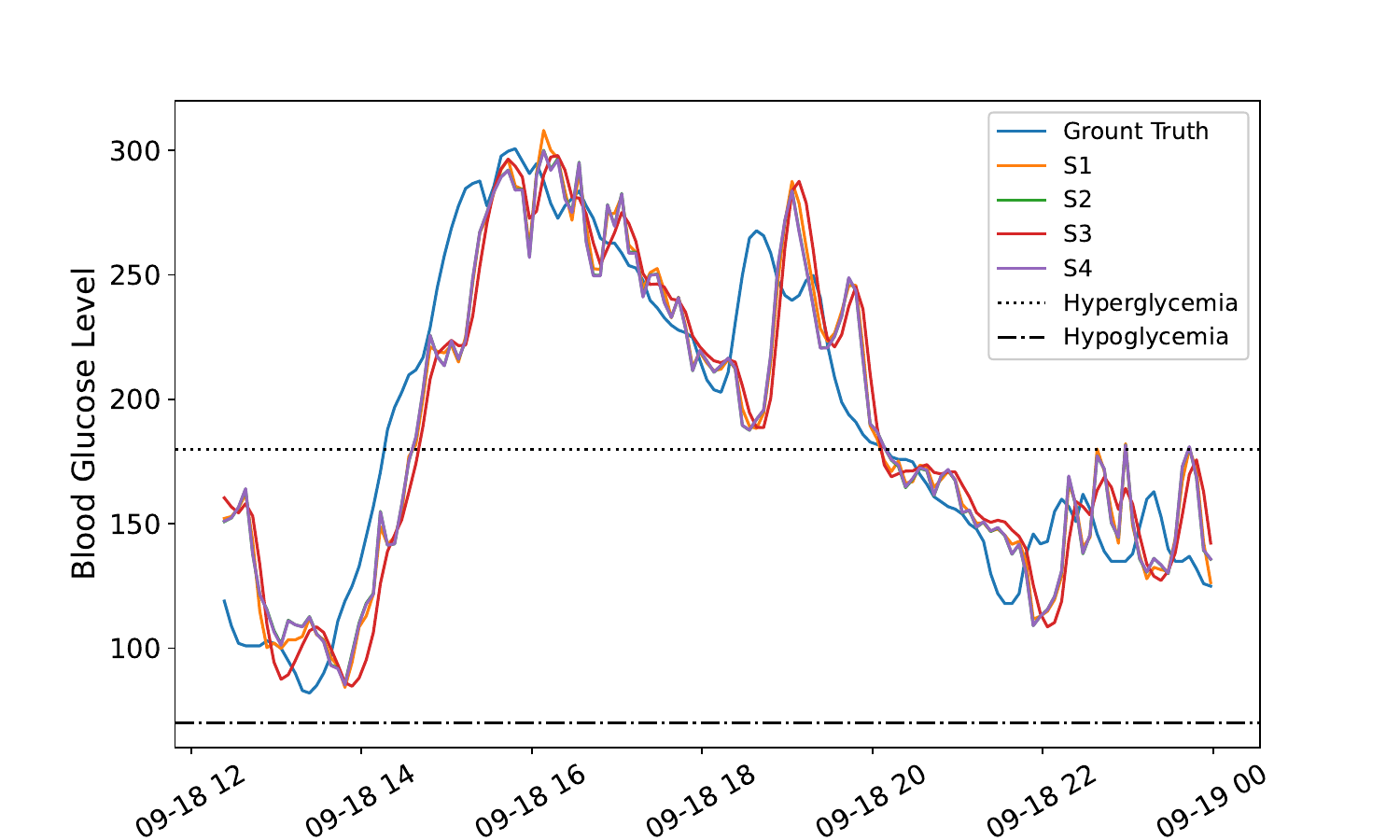}
        \caption{Stack regressor methods prediction visualization}
        \label{fig:stack_plot}
    \end{minipage}
\end{figure}

Figure \ref{fig:ml_plot}, \ref{fig:nn_plot}, \ref{fig:rl_plot}, \ref{fig:vote_plot}, and \ref{fig:stack_plot} illustrate the prediction of each method against the ground truth, plotted half a day for Subject 37. All models show a similar prediction trend as the ground truth, but a slight shift to the right.

\vspace{-25pt}
\FloatBarrier
\vspace{-10pt}
Overall, our evaluation results demonstrate the performance of various models across different glycemic conditions for blood glucose prediction.
\vspace{-10pt}

\section{Conclusion}
This research examined 15 machine-learning methods to predict blood glucose levels in patients with T1D. The methods included traditional ML, DNNs, DRL, and ensemble methods. The study focused solely on the CGM feature and ensured that the training and testing sets consisted of different subjects to avoid bias. The voting model, which combined MLP, LSTM, and GRU, displayed superior overall performance compared to other models, especially in hyperglycemic conditions. On the other hand, LSTM and the stacking model comprising MLP, LSTM, and GRU achieved the best performance in normoglycemic conditions, while DDPG alone performed better in hypoglycemic conditions. It is important to recognize the limitations of this study. The models were implemented using different libraries and frameworks, which may have led to differences in implementation and hyperparameters. Additionally, blood glucose levels are influenced by various other factors such as basal and bolus insulin dosage and meal consumption, which can impact the performance of the models. Future research work could address these limitation by focusing on extending the prediction horizon, enhancing model robustness through integration of additional physiological and contextual data, and investigating the generalizability of the models across different datasets.

\bibliography{references_diatrend}

\end{document}